How to improve the regression factor score predictor when individuals have different factor loadings

André Beauducel[*], Norbert Hilger & Anneke C. Weide

University of Bonn, Department of Psychology

**Abstract**

Previous research has shown that ignoring individual differences of factor loadings in conventional factor models may reduce the determinacy of factor score predictors. Therefore, the aim of the present study is to propose a heterogeneous regression factor score with larger determinacy than the conventional regression factor score when individuals have different factor loadings. First, a method for the estimation of individual loadings is proposed. The individual loading estimates are used to compute the heterogeneity-based regression factor score predictor. Then, a binomial test for loading heterogeneity of a factor is recommended to compute the heterogeneity-based regression factor score predictor only when the test is significant. Otherwise, the conventional regression factor score predictor should be used. A simulation study reveals that the heterogeneity-based regression factor score predictor has larger determinacy than the conventional regression factor score predictor in populations with substantial loading heterogeneity. An empirical example based on subsamples drawn randomly from a large sample of Big Five Markers indicates that the determinacy can be improved for the factor emotional stability when the heterogeneity-based regression factor score is computed.

[*]Address for correspondence:

Department of Psychology, Kaiser-Karl-Ring 9, 53111 Bonn, Germany, email: beauducel@uni-bonn.de

# Introduction

The conventional factor model consists of factor scores and factor loadings. It assumes that factor scores can differ between individuals, whereas it is implied that the loadings of the measured variables on the common factors are constant for all individuals. However, there are good reasons to expect that at least under some circumstances, individuals may have different factor loadings. Arguments for considering inter-individual loading heterogeneity have been presented from developmental psychology and behavioral genetics (Molenaar, Huizenga & Nesselroade, 2003). The meaning of individual factor loadings can be explained by an example from the concept of intelligence. In this context, factor scores represent one's intellectual capacity, whereas an individual's factor loading on a variable describes to what extent the individual makes use of intelligence for a given task. Individuals with higher loadings of the task of an intelligence factor may rely more heavily on their intelligence, whereas individuals with smaller loadings on the intelligence factor may recruit other traits to solve the task and utilize their intelligence to a smaller extent.

The assumption that an observed variable has the same factor loading for each individual of a sample is probably an over-simplification in several areas of research. This over-simplification of the factor model is, however, supported by the fact that factor analysis is typically based on the analysis of a covariance matrix of observed variables. If there are different covariances of observed variables for different individuals or for different subsamples, this cannot be detected by factor analysis of a single covariance matrix of the total sample. In line with this, Kelderman and Molenaar (2007) showed that under the assumption that normally distributed heterogenous loadings are independent of each other, the population covariance matrix of a model based on heterogenous loadings is the same as the population covariance matrix of a model based on the same loadings for all individuals. When the resulting covariance matrices of observed variables are the same for models with and without heterogenous loadings between individuals, the covariance matrices do not provide a basis for estimating heterogeneous factor loadings. Moreover, the factor scores representing the latent or "true" individual differences of the construct are indeterminate (Nicewander, 2020; Guttman, 1955) so that multiplying individual loadings with the indeterminate individual factor scores will necessarily give indeterminate results. In other words, the number of parameters of the factor model is already larger than the number of data points. Therefore, researchers rely on factor score predictors as a proxy for individual factor scores. The validity of such factor score predictors is given by their determinacy, that is, their correlation with the corresponding factor. However, given indeterminate factor scores, increasing the number of parameters by introducing individual differences on factor loadings will further increase indeterminacy. For example, one could conceive that for each individual, a different rotation of factors is possible. Moreover, Kelderman and Molenaar (2007) found that the standard likelihood-ratio goodness-of-fit statistic has little power in detecting loading heterogeneity. Therefore, considering loading heterogeneity is a challenge in the context of the exploratory factor model.

However, Ansari, Jedidi, and Dube (2002) point out that ignoring unobserved heterogeneity can lead to biased parameter estimates. They developed Markov Chain Monte Carlo (MCMC) procedures to perform Bayesian inference for confirmatory factor models with mean and covariance heterogeneity. Although this promising approach works in the context of confirmatory factor analysis, it is still relevant to consider loading heterogeneity in the context of exploratory factor analysis. As mentioned before, indeterminacy of the factor scores implies that any product of individual loadings with indeterminate factor scores will also be indeterminate. This also implies that, when a factor score predictor is specified, the product of individual loadings with the individual factor score predictor will be identified. Accordingly, Molenaar et al. (2003) investigated the effect of loading heterogeneity on the validity or determinacy of factor score predictors,, which is the correlation of the factor score predictor with the corresponding factor (Grice, 2001). They found that the determinacy coefficient reduces considerably when heterogenous loadings occur in the population model but are not specified in the factor model (Molenaar et al., 2003; Kelderman & Molenaar, 2007). Kelderman and Molenaar (2007) found that loading heterogeneity affects the distribution of the observed variables. They recommend that non-normality of the distributions of observed variables should be tested by means of the Shapiro-Wilk-Test to use factor score predictors with more confidence.

Taking these achievements as a starting point, a subsequent research question is whether the factor score predictors can be adapted to cases where heterogeneous loadings can be expected. The reduction of the factor score determinacy due to loading heterogeneity was more substantial for Bartlett's (1937) factor score predictor than for Thurstone's (1935) regression factor score predictor (Molenaar et al., 2000). Moreover, for homogeneous factor loadings, the determinacy of the regression factor score predictor (RFS) is larger than the determinacy of the Bartlett factor score predictor (Krijnen, Wansbeek, & Ten Berge, 1996), so that the RFS is considered in the following. Is it possible to minimize the loss of factor score determinacy of the RFS that cooccurs with heterogenous loadings? The present study provides a tentative solution for this problem. First, some definitions are given, second a heterogeneous regression factor score (HRFS) is proposed that may allow to minimize the loss of factor score determinacy of the RFS. Third, a simulation study is performed to compare the loss of determinacy of RFS and HRFS in models with heterogeneous factor loadings. Finally, the determinacies of RFS and HRFS are compared in an empirical dataset based on the Five Factor model, and some prospects for the use of HRFS are discussed.

**Definitions**

In a population of individuals, the common factor model (Mulaik, 2010) can be defined as

$$\mathbf{x} = \Lambda \xi + \Psi \varepsilon, \qquad (1)$$

where $\mathbf{x}$ is a vector of individual observations on $p$ observed variables, $\Lambda$ is a $p \times q$ matrix of common factor loadings, $\xi$ is a vector $q$ factor scores with the expected value $E(\xi) = 0$, the correlation matrix of common factor scores $E(\xi\xi') = \Phi$, with $diag(\Phi) = \mathbf{I}$ (variances of common factor scores), $\Psi$ is a $p \times p$ diagonal, positive definite matrix of unique factor loadings, and $\varepsilon$ is a vector of $p$ unique factor scores with $E(\varepsilon) = 0$, $E(\varepsilon\varepsilon') = \mathbf{I}$, and $E(\xi\varepsilon') = \mathbf{0}$ (uncorrelated unique factors). Accordingly, the covariance matrix of observed variables $\Sigma$ is

$$\Sigma = E(\mathbf{xx}') = \Lambda\Phi\Lambda' + \Psi^2. \qquad (2)$$

In the following, the correlation matrix of observed variables is considered, so that $diag(\Sigma) = \mathbf{I}$. The RFS, i.e., the best linear predictor (Krijnen, 2006) is defined by

$$\xi_r = \Phi\Lambda'\Sigma^{-1}\mathbf{x}. \qquad (3)$$

The correlation of the RFS with the original factor, i.e., the determinacy coefficient $\mathbf{P}$ (Grice, 2001; Guttman, 1955), can be regarded as an indicator of convergent validity of the factor score predictor. This correlation can be computed by

$$E(\xi_r\xi') = \mathbf{P} = diag(\Phi\Lambda'\Sigma^{-1}\Lambda\Phi)^{1/2}. \qquad (4)$$

For each observed variable $i$ on factor $j$, $\Lambda$ contains a common factor loading $\lambda_{ij}$, which can be conceived as the expectation of the individual common factor loadings on this variable for the population of individuals. For a finite population of $N$ individuals and completely independent individual loadings (Klederman & Molenaar, 2007), the expected value of the individual loadings is

$$\lambda_{ij} = E(\lambda_{ijk}) = \frac{1}{N}\sum_{k}^{N} \lambda_{ijk}. \tag{5}$$

For $\Phi = I$ and $-1 > \lambda_{ij} < 1$, it is possible that there are individual differences of factor loadings, so that the variance of factor loadings across individuals is greater zero:

$$\sigma_{ij}^2 = \frac{1}{N}\sum_{k}^{N}(\lambda_{ijk} - \lambda_{ij})^2 > 0. \tag{6}$$

For $\Phi \neq I$, individual differences may also occur for $|\lambda_{ij}| \geq 1$, and individual differences of factor inter-correlations may also occur. However, the estimation of individual factor loadings $\lambda_{ijk}$ is already a challenge for the orthogonal factor models, so that the present approach is limited to these models.

**Estimation of individual factor loadings**

The proposed estimation procedure is described for a sample of $n$ individuals and for an orthogonal factor model. First, factor analysis of $q$ factors and $p$ variables for the total sample yields

$$\widehat{\Sigma} = \widehat{\Lambda}\widehat{\Lambda}' + \widehat{\Psi}^2. \tag{7}$$

Model misfit occurs in the sample, so that the sample covariance matrix differs from the population covariances, $E(\mathbf{xx}') = \mathbf{S} \neq \widehat{\Sigma}$. In consequence, the determinacy coefficient, that is, the correlation of $\hat{\xi}_r$ with $\xi$ is

$$E\left(diag(\hat{\xi}_r\hat{\xi}_r')^{-1/2}\hat{\xi}_r\xi'\right) = \mathbf{P} = diag\left(\widehat{\Lambda}'\widehat{\Sigma}^{-1}\mathbf{S}\widehat{\Sigma}^{-1}\widehat{\Lambda}\right)^{-1/2} diag\left(\widehat{\Lambda}'\widehat{\Sigma}^{-1}\widehat{\Lambda}\right). \tag{8}$$

Similar to Cook's (1977) ideas on the influence of a single individual on linear regression results, the determinacy of the RFS $\hat{\xi}_{r(-k)}$ is computed from factor analysis when the data of the $k$th individuum are deleted. Factor analysis of $\mathbf{S}_{(-k)}$, the sample covariance matrix based on one eliminated case, yields the loading estimates $\widehat{\Lambda}_{(-k)}$, where $(-k)$ refers to the parameters estimated without the data of the individual $k$. If $q > 1$, it is recommended to perform orthogonal target rotation of $\widehat{\Lambda}_{(-k)}$ towards $\widehat{\Lambda}$ (Schoenemann, 1966) to minimize effects of different positions of the factor axes of $\widehat{\Lambda}_{(-k)}$ and $\widehat{\Lambda}$ on results. The determinacy coefficient for the factors when the data of the $k$th individuum are deleted, is

$$E\left(diag\left(\hat{\xi}_{r(-k)}\hat{\xi}'_{r(-k)}\right)^{-1/2}\hat{\xi}_{r(-k)}\xi'_{(-k)}\right) = \mathbf{P}_{(-k)}$$

$$= diag\left(\widehat{\Lambda}'_{(-k)}\widehat{\Sigma}^{-1}_{(-k)}\mathbf{S}_{(-k)}\widehat{\Sigma}^{-1}_{(-k)}\widehat{\Lambda}_{(-k)}\right)^{-1/2} diag(\widehat{\Lambda}'_{(-k)}\widehat{\Sigma}^{-1}_{(-k)}\widehat{\Lambda}). \tag{9}$$

The effect of the $k$th individuum on the determinacies of the factor score predictors is

$$\Delta \mathbf{P}_k^2 = \mathbf{P}^2 - \mathbf{P}^2_{(-k)}. \tag{10}$$

Positive values of $\Delta P_k^2$ indicate that the *k*th individuum contributes positively to the determinacy of the factor score predictor. The effect of the *k*th individuum on the RFS for a single factor is

$$\Delta \rho_{jk}^2 = \rho_j^2 - \rho_{j(-k)}^2. \tag{11}$$

Although $\Delta \rho_{jk}^2$ allows to identify individuals for whom the prediction of the common factor is above or below average, it does not help to compensate for the effect of low prediction. To compensate for the different influence of individuals on determinacy, it is proposed to estimate the size of the individual loadings $\hat{\lambda}_{ijk}$. If a loading decreases when the data of individual *k* are deleted, the individual *k* has an increasing effect on the overall loading. Therefore, the effect of the data of an individuum *k* on a loading can be estimated by

$$\Delta \hat{\lambda}_{ijk} = \left| \hat{\lambda}_{ij}^2 \, sgn(\hat{\lambda}_{ij}) - \hat{\lambda}_{ij(-k)}^2 sgn(\hat{\lambda}_{ij(-k)}) \right|^{1/2} sgn\left( \hat{\lambda}_{ij}^2 \, sgn(\hat{\lambda}_{ij}) - \hat{\lambda}_{ij(-k)}^2 sgn(\hat{\lambda}_{ij(-k)}) \right), \tag{12}$$

where "*sgn*" is the sign-function. The multiplication with the sign of the loadings maintains their sign after being squared. As $\hat{\lambda}_{ij}^2$ is based on *n* individuals and $\hat{\lambda}_{ij(-k)}^2$ is based on $n - 1$ individuals, the estimated effect of an individuum *k* on $\hat{\lambda}_{ij}^2$ will decrease with larger *n*. However, the effect of sampling error on $\hat{\lambda}_{ij}^2$ and $\hat{\lambda}_{ij(-k)}^2$ also decreases with *n*, so that the estimation of $\hat{\lambda}_{ij(-k)}^2$ is not necessarily worse for larger *n*. Moreover, individual differences of measurement error may also affect $\hat{\lambda}_{ij(-k)}^2$. For this reason, the optimal calibration of $\hat{\lambda}_{ij(-k)}^2$ is unknown. In order to keep the effect of measurement error in $\hat{\lambda}_{ij(-k)}^2$ on $\Delta \hat{\lambda}_{ijk}$ and thereby on $\hat{\lambda}_{ijk}$ small, it is proposed to estimate $\hat{\lambda}_{ijk}$ from a weighted aggregate of $\Delta \hat{\lambda}_{ijk}$ and $\hat{\lambda}_{ij}$. Effects of $\Delta \hat{\lambda}_{ijk}$ on individual loadings should be more substantial for larger than for smaller total sample absolute loadings. Accordingly, $\Delta \hat{\lambda}_{ijk}$ is weighted by $w = |\hat{\lambda}_{ij}|/mean(|\hat{\lambda}_j|)$ so that

$$\hat{\lambda}_{ijk} = \hat{\lambda}_{ij} + w \Delta \hat{\lambda}_{ijk}. \tag{13}$$

In a stepwise procedure, $w \Delta \hat{\lambda}_{ijk}$ is added to only one element $\hat{\lambda}_{ij}$ resulting in $\widehat{\Lambda}_{ijk}$, in which only element *ij* differs from the corresponding element in $\widehat{\Lambda}$. In order to minimize the effect of sampling error on $\hat{\lambda}_{ijk}$, we propose to add $w \Delta \hat{\lambda}_{ijk}$ to $\hat{\lambda}_{ij}$ only when the squared difference of the non-diagonal elements of the correlation matrix reproduced from $\widehat{\Lambda}_{ijk}$ and the non-diagonal elements of $\mathbf{S}_{(-k)}$ is larger than the squared difference of the non-diagonal elements of the correlation matrix reproduced from the loadings $\widehat{\Lambda}$ and the non-diagonal elements of $\mathbf{S}$. Accordingly, the individual loadings are estimated as follows:

$$\tilde{\hat{\lambda}}_{ijk} = \begin{cases} \hat{\lambda}_{ij} + w \Delta \hat{\lambda}_{ijk} \text{ if } SSQ\left( \widehat{\Lambda}_{ijk} \widehat{\Lambda}'_{ijk} - diag\left( \widehat{\Lambda}_{ijk} \widehat{\Lambda}'_{ijk} \right) - \left( \mathbf{S}_{(-k)} - diag(\mathbf{S}_{(-k)}) \right) \right) \\ \qquad\qquad > SSQ\left( \widehat{\Lambda}\widehat{\Lambda}' - diag\left( \widehat{\Lambda}\widehat{\Lambda}' \right) - (\mathbf{S} - diag(\mathbf{S})) \right) \\ \hat{\lambda}_{ij} \text{ else} \end{cases}, \tag{14}$$

where "*SSQ*" denotes the sum of squares. Heywood cases may occur more often for $\tilde{\hat{\lambda}}_{ijk}$ than for $\hat{\lambda}_{ij}$, and it should be considered to reset the respective individual loadings to more realistic absolute loadings (e.g., .99).

**The heterogeneity-based regression factor score predictor**

Making use of the estimated individual loadings, the heterogeneity-based regression factor score predictor (HRFS) is then computed for each individual as

$$\hat{\xi}_{rk} = diag\left(\widetilde{\tilde{\Lambda}}'_k \hat{\Sigma}^{-1} S \hat{\Sigma}^{-1} \widetilde{\tilde{\Lambda}}_k\right)^{-1/2} \widetilde{\tilde{\Lambda}}'_k \hat{\Sigma}^{-1} x. \quad (15)$$

The estimated correlation of $\hat{\xi}_{rk}$ with $\xi$, the individual determinacy of the HRFS is

$$E\left(diag(\hat{\xi}_{rk}\hat{\xi}'_{rk})^{-1/2} \hat{\xi}_{rk}\xi'\right) = \hat{P}_k = Mean_k\left(diag\left(\widetilde{\tilde{\Lambda}}'_k \hat{\Sigma}^{-1} S \hat{\Sigma}^{-1} \widetilde{\tilde{\Lambda}}_k\right)^{-1/2} diag\left(\widetilde{\tilde{\Lambda}}'_k \hat{\Sigma}^{-1} \widetilde{\tilde{\Lambda}}_k\right)\right), (16)$$

Where "$Mean_k$" denotes the mean across all individuals in a sample. It is proposed to estimate $\hat{P}_k$ and the HRFS only when loadings are heterogeneous between individuals. Otherwise, the estimation of $\hat{P}$ and the conventional RFS is recommended. Although deviations from normality may be an indicator of loading heterogeneity (Kelderman & Molenaar, 2007), several other effects may cause non-normal distributions. Therefore, a more specific indicator of loading heterogeneity is proposed.

**Estimation of loading heterogeneity**

Loading heterogeneity can be assessed when, in a sample of $n$ individuals, $n$ factor analyses with the data minus one individuum $k$ are performed. For each loading on each factor, the inter-individual standard deviation $\hat{\sigma}(\hat{\lambda}_{ij(-k)})$ can be compared with the inter-individual standard deviation of loadings of $n$ factor analyses based on a sample of $n$ cases minus one case, drawn from a simulated population without inter-individual variability of loadings, that is, $\lambda_{ijk} = \lambda_{ij}$, for all $k$. This standard deviation is based on the loadings $\hat{\lambda}_{ij(-k)}$, resulting directly from leaving individual $k$ out, as used in Equation 12 (it is not the individual loading $\tilde{\tilde{\lambda}}_{ijk}$) because the aim is to estimate whether leaving successively one individual out results in substantial loading heterogeneity. Let $\hat{\sigma}(\hat{\lambda}_{0,ij(-k)})$ denote the standard deviation of loadings based on $n$ factor analyses with a sample of $n – 1$ individuals drawn randomly from a simulated population with zero loading heterogeneity. The effect of leaving one individual out on loading heterogeneity may depend on loading magnitude in the total sample because there might be ceiling effects. Therefore, the population loading of each simulated variable should equal the mean of the individual loadings of the corresponding variable in the empirical data set, that is,

$$\lambda_{0,ij} = Mean(\hat{\lambda}_{ij(-k)}). \quad (17)$$

In the simulated population, $\lambda_{0,ij}$ is constant for all $k$ individuals, so that $\sigma(\lambda_{0,ijk}) = 0$. In consequence, any $\hat{\sigma}(\hat{\lambda}_{0,ij(-k)}) > 0$, found in a sample drawn from this population is due to sampling error. If $n_d$ samples are drawn from this population, the mean of the resulting standard deviations of loadings can be compared with the empirical standard deviation of loadings for the corresponding variable $i$ on factor $j$. If $Mean_{n_d}\left(\hat{\sigma}(\hat{\lambda}_{0,ij(-k)})\right) < \hat{\sigma}(\hat{\lambda}_{ij(-k)})$, one would conclude that the empirical loading heterogeneity is not due to sampling error.

To obtain an indicator for the loading heterogeneity of a factor, the number of variables with loading heterogeneity greater than sampling error is counted for the respective factor. The following index is one if loading heterogeneity of variable $i$ occurs, that is, if $\hat{\sigma}(\hat{\lambda}_{ij(-k)})$ is greater than $Mean_{n_d}\left(\hat{\sigma}(\hat{\lambda}_{0,ij(-k)})\right)$:

$$\vartheta_{ij} = \begin{cases} 1 \text{ if } Mean_{n_d}\left(\hat{\sigma}(\hat{\lambda}_{0,ij(-k)})\right) < \hat{\sigma}(\hat{\lambda}_{ij(-k)}) \\ 0 \text{ if } Mean_{n_d}\left(\hat{\sigma}(\hat{\lambda}_{0,ij(-k)})\right) \geq \hat{\sigma}(\hat{\lambda}_{ij(-k)}) \end{cases}. \tag{18}$$

The number $\vartheta_j = \sum_{i=1}^{p_j} \vartheta_{ij}$ for each factor follows a binomial distribution where "success" is defined by $\vartheta_{ij} = 1$, the number of trials by $p$ (number of variables), and the success probability is .50. The minimum number of successes that $\vartheta_j$ should reach to reject the assumption of homogeneity is $\vartheta_{crit}$. The decision on the hypothesis of homogeneity can therefore be made with a right-tailed binomial test, whereby the problem arises that with only a few variables per factor, only a few exact significance levels can be determined. Using the critical values listed in Table 1, a significance level of $\alpha \leq .25$ is ensured. The assumption of homogeneity is therefore rejected if $\vartheta_j \geq \vartheta_{crit}$, so that HRFS should be computed, whereas the conventional RFS should be computed if $\vartheta_j < \vartheta_{crit}$.

Table 1. Cut-off values $\vartheta_{crit}$ for the index of loading heterogeneity for $p \leq 12$ (number of variables) resulting in $\alpha \leq .25$

| $p$ | $\vartheta_{crit}$ | $\alpha_{exact}$ |
|---|---|---|
| 2 | 2 | .2500 |
| 3 | 3 | .1250 |
| 4 | 3 | .2500 |
| 4 | 4 | .0625 |
| 5 | 4 | .1875 |
| 6 | 5 | .1094 |
| 7 | 5 | .2266 |
| 8 | 6 | .1445 |
| 9 | 7 | .0898 |
| 10 | 7 | .1719 |
| 11 | 8 | .1133 |
| 12 | 8 | .1938 |
| 12 | 9 | .0730 |

For $p \leq 3$, it is necessary to use a cut-off value of $\vartheta_{crit} = p$. Lower cut-off values with $\vartheta_{crit} < p$ can only be used for $p > 3$, and a more fine-grained control of $\alpha$ is only possible for large $p$. However, an advantage of this indicator of loading heterogeneity is that it can also be used for categorical factor analysis. Therefore, the proposed index yields not necessarily the same results as the Shapiro-Wilk-test for non-normality.

**Simulation Study**

*Conditions and specification*

A simulation study was performed to investigate whether the HRFS has a larger validity, that is, determinacy coefficient than the conventional RFS when loading heterogeneity occurs. In an empirical study, the determinacy coefficient $\hat{\rho}_r$ can only be estimated by Equation 8, and $\hat{\rho}_{rk}$ can only be estimated by Equation 16. In contrast, in a simulation study based on generated factor scores ξ, these correlations can be computed directly and averaged across samples. Therefore, the parameter-based determinacy estimates $\hat{\rho}_r$ and $\hat{\rho}_{rk}$, can be compared with the factor score based determinacy estimated by $\hat{\rho}_{\xi r} = E(\hat{\xi}_r \xi')$, and $\hat{\rho}_{\xi rk} = E(\hat{\xi}_{rk} \xi')$. As mentioned above, $\hat{\rho}_{\xi rk}$ and HRFS should only be estimated when there is loading heterogeneity, otherwise $\hat{\rho}_{\xi r}$ and RFS are considered more appropriate. Therefore, the dependent variables of the simulation study are defined as follows. RFS $\hat{\xi}_r$ was computed for each sample in each condition as a basis for $\hat{\rho}_{\xi r}$. HRFS was only used for the computation of $\hat{\rho}_{\xi rk}$ when $\vartheta_j \geq \vartheta_{crit}$, so that

$$\tilde{\rho}_{\xi rk} = \begin{cases} \hat{\rho}_{\xi rk} & if \ \vartheta_j \geq \vartheta_{crit} \\ \hat{\rho}_{\xi r} & if \ \vartheta_j < \vartheta_{crit} \end{cases}, \quad (19)$$

and

$$\tilde{\rho}_{rk} = \begin{cases} \hat{\rho}_{rk} & if \ \vartheta_j \geq \vartheta_{crit} \\ \hat{\rho}_r & if \ \vartheta_j < \vartheta_{crit} \end{cases}. \quad (20)$$

A continuous proportion of $\vartheta_{crit}/p = 0.83$ was used for the conditions of the simulation study, so that the minimum of variables with loading heterogeneity defining a heterogenous factor remained constant across conditions. In consequence, the probability α for the rejection of the hypothesis that a factor has homogenous loadings across individuals decreased with *p*. For *q* = 1 and *p* = 6, $\vartheta_{crit}$ = 5 was used, resulting in $\alpha_{exact}$ = .109 and $\vartheta_{crit}$ = 9 for *p* = 12, resulting in $\alpha_{exact}$ = .073 (see Table 1). For *q* = 3 and *p* = 18, $\vartheta_{crit}$ = 15 was used, resulting in $\alpha_{exact}$ = .004 and for *p* = 36, $\vartheta_{crit}$ = 30 was used, resulting in $\alpha_{exact}$ < .001. Thus, the condition for the computation of HRFS became more conservative with increasing *p*. Note that more similar rejection rates of the loading homogeneity hypothesis would have implied that the proportion of variables with loading heterogeneity differs across the conditions of the simulation study.

Independent variables of the simulation study were the loading heterogeneity with σ($\Lambda_{pop}$) ∈ {.00, .25, .50, .75}, $p \in \{6, 12\}, q \in \{1, 3\}$, and mean salient loading size μ($\Lambda_{pop}$) ∈ { .60, .70}, and sample size *n* ∈ {150, 600}. For σ($\Lambda_{pop}$), each sample of loadings was fixed to have the given population standard deviation. If absolute loadings > .98 occurred, all population loadings of a factor were divided by a constant so that the maximum absolute loading of a factor on a variable for each individual was .98, which slightly reduced σ($\Lambda_{pop}$), especially in the *n* = 150 and σ($\Lambda_{pop}$) = .75 conditions. Note that the means and standard deviations of the loadings were fixed conditions, that is, each condition had exactly the specified mean and standard deviation of loadings. This yielded 4 × 2 × 2 × 2 × 2 = 64 conditions. For each condition, 1,000 samples were drawn from the population.

For each sample of each condition, $n \times n_d$ factor analyses for the computation of $\vartheta_j$ were performed. In order to keep the simulation study feasible, we investigated $n_d$ = 50 simulated samples. Accordingly, for each condition based on $n$ = 150 this led to 150 × 50 × 1,000 = 7.5e6 factor analyses and for each condition based on $n$ = 600 this led to 30e6 factor analyses. Overall, 37.5e6 factor analyses were performed for the conditions listed above, resulting in about one month of simulation time when about 14 factor analyses per second were performed simultaneously. It was therefore not considered to investigate a larger set of conditions here.

As mentioned above, the effect of factor rotation on results for $q$ = 3 was minimized by means of orthogonal target rotation according to Schoenemann (1966) towards the target matrix of salient loadings in the total samples. Moreover, $\hat{\Lambda}_{(-k)}$, the loadings resulting from $n$ factor analyses of the sample minus one eliminated case, were rotated towards $\hat{\Lambda}$, the loadings of the total sample analysis, by means of orthogonal target rotation. Thereby, the factor loadings $\hat{\lambda}^2_{ij(-k)}$ were as similar as possible to $\hat{\lambda}^2_{ij}$, the factor loadings in the total sample. The simulation was performed with R4.4.1, R-packages (MASS, psych, matrixStats.

*Results*

The effect of loading heterogeneity on determinacy estimates was investigated by means of a repeated-measures ANOVA comprising parameter-based versus factor-score based determinacy coefficients (PAR-SCO) and the RFS-based versus HRFS-based determinacy coefficients (RFS-HRFS) as within-subjects factors and the conditions, $q$, $p$, $\mu(\Lambda_{pop})$, $\sigma(\Lambda_{pop})$, and $n$ as between-subjects factors. The condition with $n$ = 150 and $\sigma(\Lambda_{pop})$ > .75 was excluded because only about 10% of the individual factor analyses converged in this condition. Significance levels were not considered because due to the large sample size (1,000 cases per cell), even effects with $\eta^2_p$ = .001 were significant at an alpha-level of .001. The PAR-SCO main effect was large ($\eta^2_p$ = .26) although -across all conditions- the mean of the score-based determinacy coefficients ($M$ = .91, $SE$ < .001) was only slightly larger than the mean of the parameter-based determinacy coefficients ($M$ = .89, $SE$ < .001). The RFS-HRFS main effect was even larger ($\eta^2_p$ = .73) although the mean determinacy coefficients for the HRFS were again only slightly larger ($M$ = .91, $SE$ < .001) than the mean determinacy coefficients of the RFS ($M$ = .89, $SE$ < .001). The effect size of the PAR-SCO × $\sigma(\Lambda_{pop})$ interaction ($\eta^2_p$ = .33) and the effect size of the RFS-HRFS × $\sigma(\Lambda_{pop})$ interaction ($\eta^2_p$ = .56) were large indicating that loading heterogeneity affects the difference between parameter-based and score-based determinacy coefficients as well as the difference of determinacies between HRFS and RFS (see Figures 1 and 2). For $\sigma(\Lambda_{pop})$ = .00, the mean determinacy was $M$ = .93 ($SE$ < .001) for RFS and $M$ = .93 ($SE$ < .001) for HRFS. For $\sigma(\Lambda_{pop})$ = .75 the mean determinacy was $M$ = .85 ($SE$ < .001) for RFS and $M$ = .88 ($SE$ < .001) for HRFS. Besides these effects, which were most interesting for the comparison between RFS and HRFS, there were very large main effects of $p$ ($\eta^2_p$ = .90) and $\mu(\Lambda_{pop})$ ($\eta^2_p$ = .73). These main effects were as expected with larger mean determinacy for $p$ = 12 ($M$ = .93, $SE$ < .001) than for $p$ = 6 ($M$ = .87, $SE$ < .001) and for $\mu(\Lambda_{pop})$ = .70 ($M$ = .92, $SE$ < .001) than for $\mu(\Lambda_{pop})$ = .60 ($M$ = .88, $SE$ < .001). For $\sigma(\Lambda_{pop})$, the main effect was very large ($\eta^2_p$ = .84), with ($M$ = .93, $SE$ < .001) for $\sigma(\Lambda_{pop})$ = .00 and ($M$ = .87, $SE$ < .001) for $\sigma(\Lambda_{pop})$ = .75. Compared to these effects, the main effect for $q$ was smaller ($\eta^2_p$ = .17) with $M$ = .91 ($SE$

< .001) for $q = 3$ and $M = .89$ ($SE < .001$) for $q = 1$. The main effect of $n$ was small ($\eta_p^2 = .06$) with $M = .90$ ($SE < .001$), both for $n = 150$ and for $n = 600$ (as mean differences occurred on the third decimal place).

To investigate whether these results could be replicated when the initial total-sample solution was based on Varimax-rotated factors, we performed the analysis for $q = 3$ and $p/q = 6$, for $n = 150$ and $n = 600$. The results for Varimax-rotated solutions (see Figure 3) were similar to the results for the corresponding target-rotated solutions (see Figure 2).

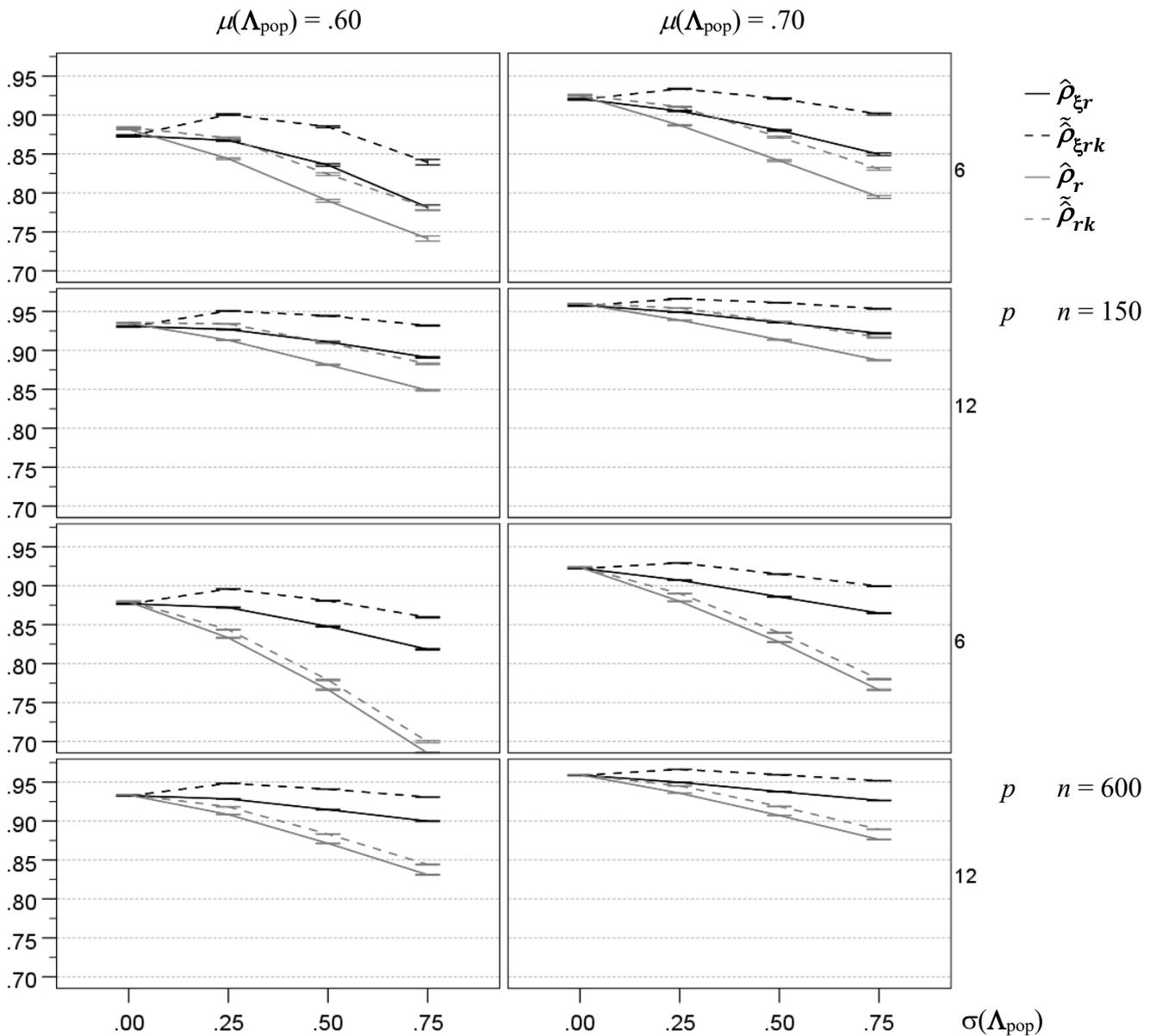

Figure 1. Mean determinacy coefficients based on model parameters and RFS ($\hat{\rho}_r$), on model parameters and HRFS ($\tilde{\hat{\rho}}_{rk}$), on scores and RFS ($\hat{\rho}_{\xi r}$), and on scores and HRFS ($\tilde{\hat{\rho}}_{\xi rk}$) for a single factor ($q = 1$), $p = 6$, and $p = 12$ variables; $\mu(\Lambda_{pop})$ is the expected value of the population loadings; $\sigma(\Lambda_{pop})$ is the expected value of loading heterogeneity; the error bars mark the standard errors.

As the parameter-based determinacies may be used as estimates for the score-based determinacies in empirical settings (when score-based determinacies cannot be computed), the effect of loading heterogeneity on parameter-based and score-based determinacies was compared. To compare the effect of loading heterogeneity on parameter-based and score-based determinacies, the effect of σ($\Lambda_{pop}$) on $\Delta\tilde{\hat{\rho}}_{\xi rk-\xi r} = \tilde{\hat{\rho}}_{\xi rk} - \hat{\rho}_{\xi r}$, the difference of the score-based determinacies of HRFS and RFS and on $\Delta\tilde{\hat{\rho}}_{rk-r} = \tilde{\hat{\rho}}_{rk} - \hat{\rho}_{r}$, the difference of the parameter-based determinacies of HRFS and RFS were compared (see Figure 4).

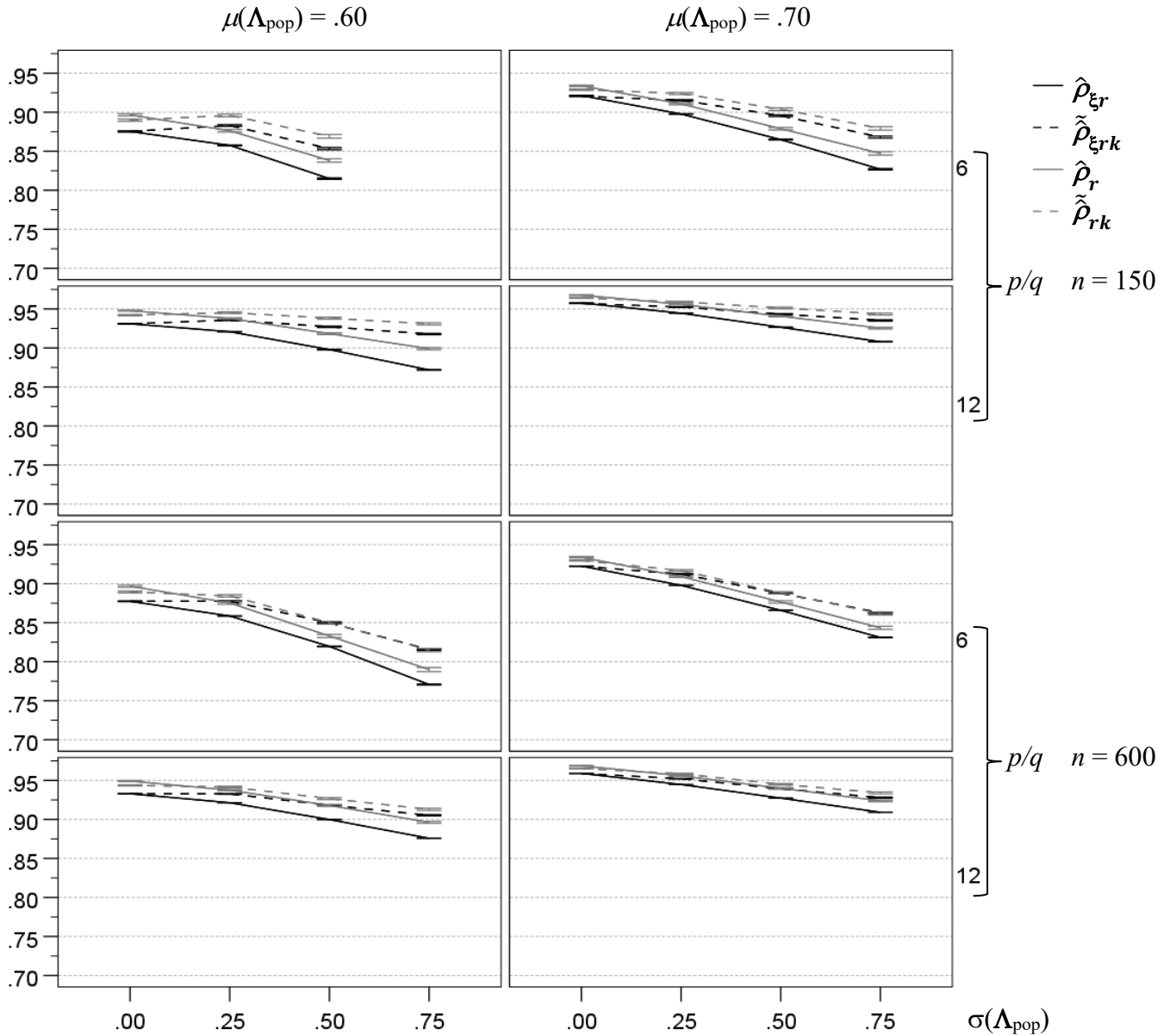

Figure 2. Mean determinacy coefficients based on model parameters and RFS ($\hat{\rho}_r$), on model parameters and HRFS ($\tilde{\hat{\rho}}_{rk}$), on scores and RFS ($\hat{\rho}_{\xi r}$), and on scores and HRFS ($\tilde{\hat{\rho}}_{\xi rk}$) for three factors ($q = 3$), $p/q = 6$, and $p/q = 12$ variables per factor; $\mu(\Lambda_{pop})$ is the expected value of the population loadings; σ($\Lambda_{pop}$) is the expected value of loading heterogeneity; the error bars mark the standard errors; for $q = 3$, $n = 150$ and σ($\Lambda_{pop}$) > .75, only about 10% of the individual factor analyses converged, so that no means were computed for this condition.

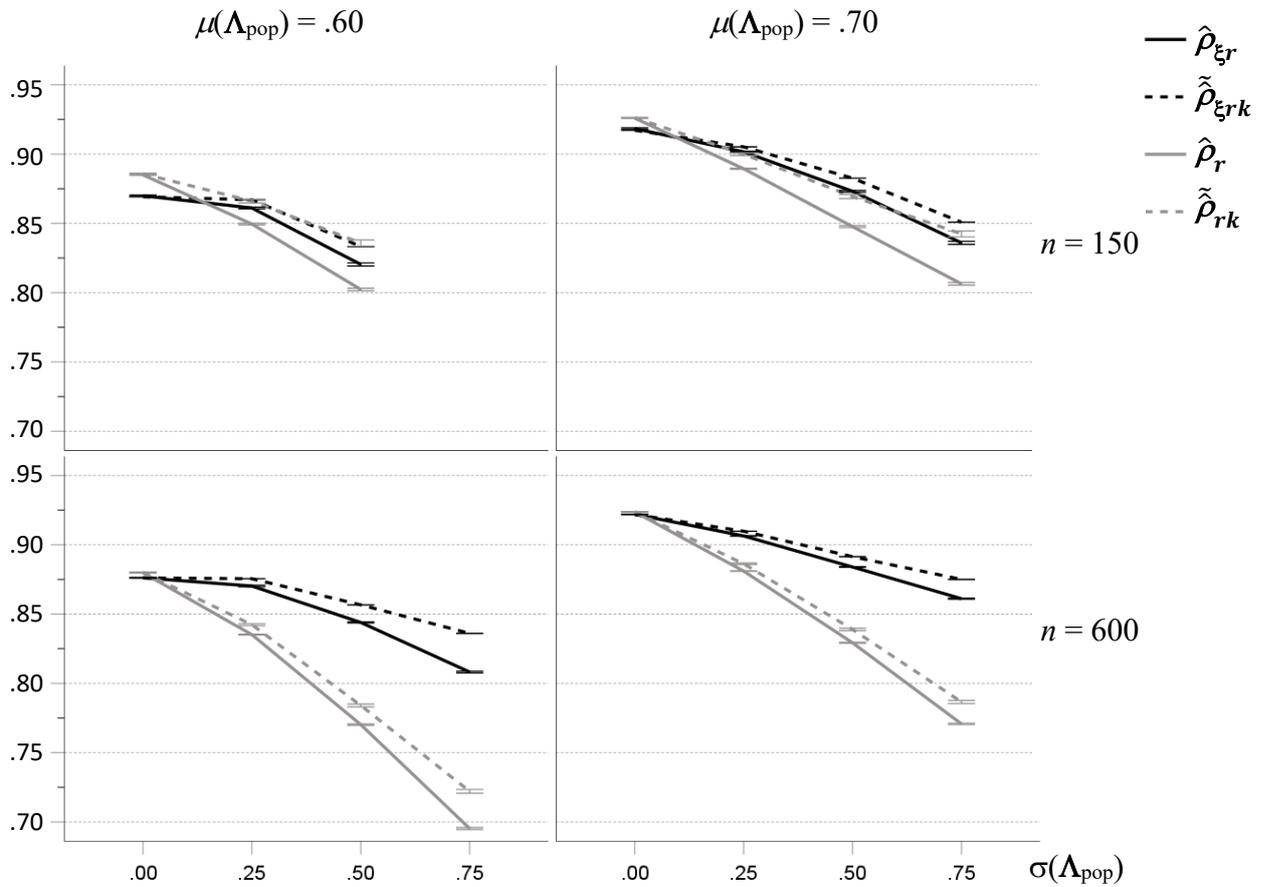

Figure 3. Mean determinacy coefficients based on model parameters and RFS ($\hat{\rho}_r$), on model parameters and HRFS ($\tilde{\hat{\rho}}_{rk}$), on scores and RFS ($\hat{\rho}_{\xi r}$), and on scores and HRFS ($\tilde{\hat{\rho}}_{\xi rk}$) for three factors ($q = 3$), $p/q = 6$, and $p/q = 12$ variables per factor for Varimax-rotated solutions; $\mu(\Lambda_{pop})$ is the expected value of the population loadings; $\sigma(\Lambda_{pop})$ is the expected value of loading heterogeneity; the error bars mark the standard errors. For $q = 3$, $n = 150$ and $\sigma(\Lambda_{pop}) > .75$, only about 10% of the individual factor analyses converged, so that no means were computed for this condition.

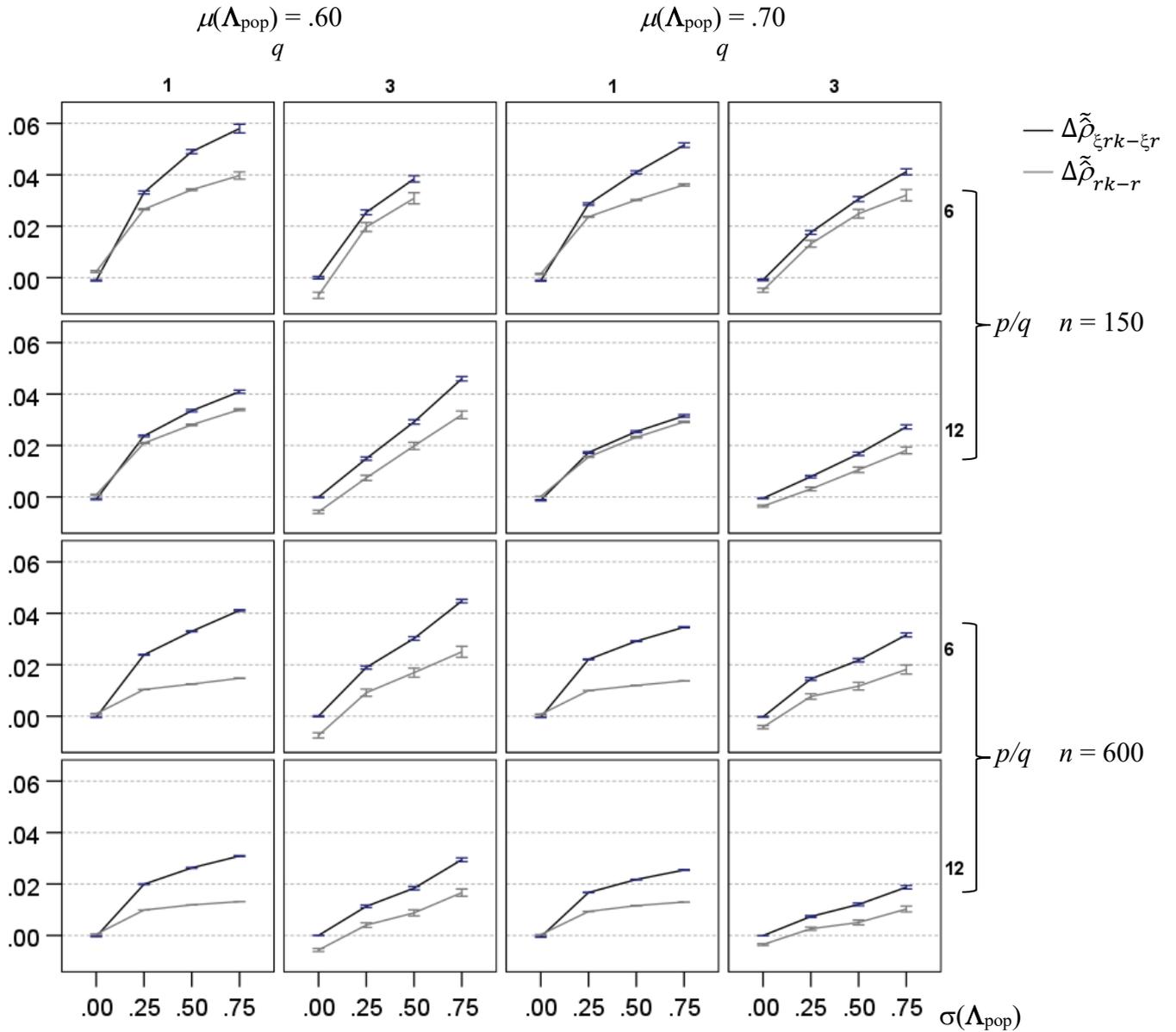

Figure 4. The means of $\Delta\tilde{\hat{\rho}}_{\xi rk-\xi r} = \tilde{\hat{\rho}}_{\xi rk} - \hat{\rho}_{\xi r}$ and $\Delta\tilde{\hat{\rho}}_{rk-r} = \tilde{\hat{\rho}}_{rk} - \hat{\rho}_{r}$ plotted for $p$ (number of variables), $q$ (number of factors), $\mu(\Lambda_{\text{pop}})$, and $\sigma(\Lambda_{\text{pop}})$ as mean and standard deviation of population loadings; the error bars mark the standard errors.

Overall, the means of $\Delta\tilde{\hat{\rho}}_{rk-r}$ were smaller than the means of $\Delta\tilde{\hat{\rho}}_{\xi rk-\xi r}$. However, the means of $\Delta\tilde{\hat{\rho}}_{rk-r}$ and $\Delta\tilde{\hat{\rho}}_{\xi rk-\xi r}$ both increase with $\sigma(\Lambda_{\text{pop}})$. Accordingly, $\Delta\tilde{\hat{\rho}}_{rk-r}$ can be used as a lower-bound estimate of $\Delta\tilde{\hat{\rho}}_{\xi rk-\xi r}$, i.e., when $\tilde{\hat{\rho}}_{rk}$ is greater than $\hat{\rho}_{r}$, there are good reasons to expect that $\tilde{\hat{\rho}}_{\xi rk}$ is greater than $\hat{\rho}_{\xi r}$.

**Empirical example**

The empirical example dataset was based on answers to the 50 IPIP Big Five Factor Markers (Goldberg, 1992), updated at 11/08/2018 and retrieved at 13/11/2024 from https://openpsychometrics.org/_rawdata/. The five factors are Extraversion (E), Agreeableness (A), Conscientiousness (C), Emotional Stability (ES), and Intellect/Imagination (I). Each factor was measured by means of 10 items with five response categories, the direction of item-scoring was altered. The dataset was collected from 2016 to 2018 through an interactive on-line personality test and contained 1,015,342 cases. However, it is recommended in the codebook to use only cases with a single user IP. Accordingly, only 696,854 cases with a single user IP were used. No demographic information was available. 100 random subsamples with $n = 150$, $n = 600$, and $n = 1,000$ cases were drawn with replacement from the total sample. In a first step, principal axis factor analysis with $q = 5$ and subsequent orthogonal target-rotation towards the intended five factor loading pattern was performed as a basis for the computation of $\hat{\rho}_r$. In a second step, the analyses were performed for factors that were Varimax-rotated in the total sample. This allowed to investigate the effect of analytic rotation on results. A significance level of α ≤ .20 corresponding to $\vartheta_{crit} = 29$ and α*exact* = .16 was used for the assessment of loading heterogeneity and the computation of $\hat{\tilde{\rho}}_{rk}$.

The results for the factors based on orthogonal target-rotation in the total sample indicated that $\hat{\tilde{\rho}}_{rk}$ was larger than $\hat{\rho}_r$ for the ES factor in samples of $n = 600$ and $n = 1,000$ because the 95% confidence intervals did not overlap (see Figure 5). Thus, for ES, HRFS had on average larger determinacies than RFS. For the other Big Five factors, there were no significant differences in determinacy coefficients between HRFS and RFS. The same result was found for Varimax-rotated factors (see Figure 6). Hence, the effect of loading heterogeneity on determinacy and the possible advantage of HRFS over RFS did not only occur for target-rotated factors.

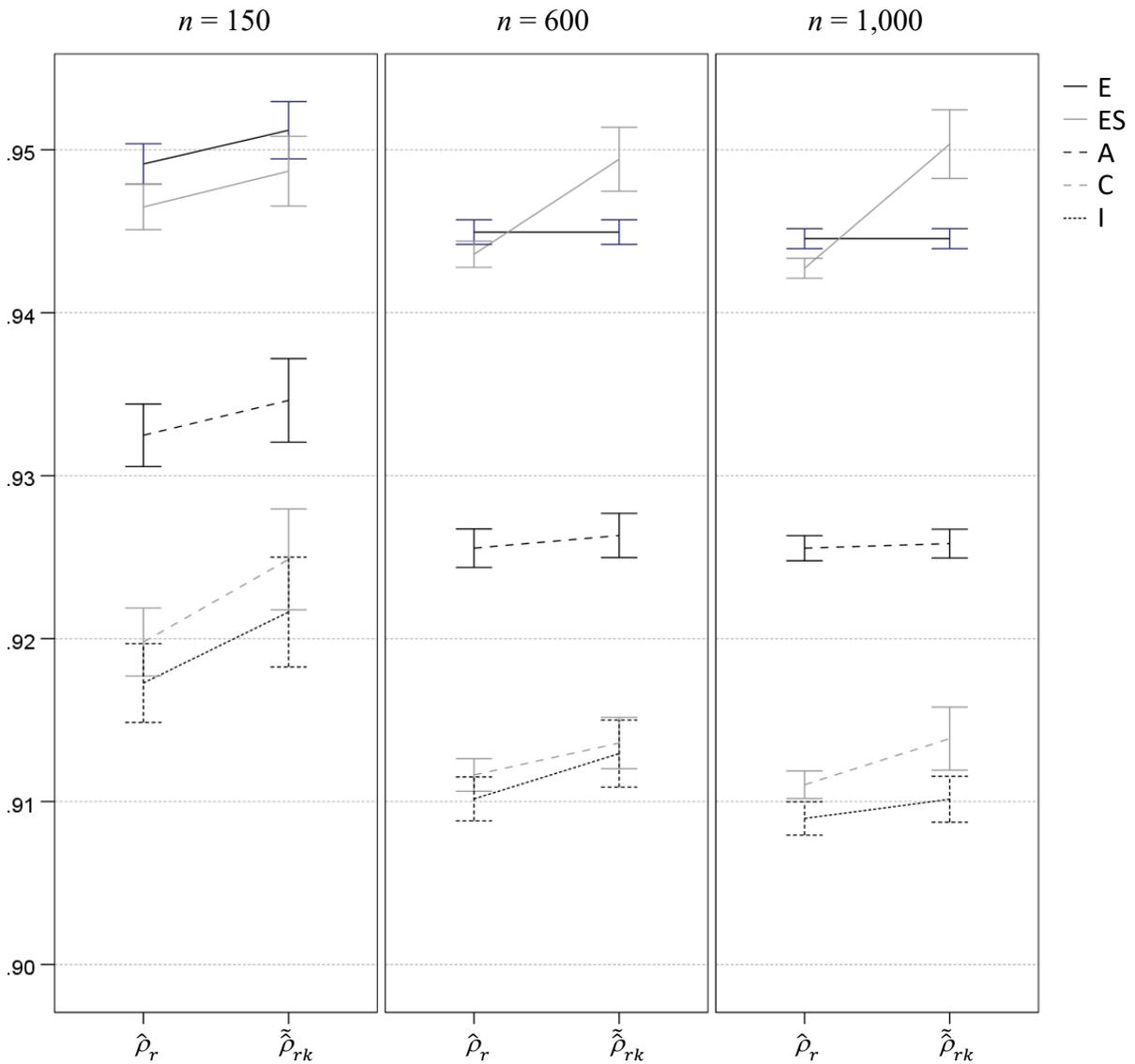

Figure 5. Mean parameter-based determinacy of regression factor scores (RFS) $\hat{\rho}_r$ and heterogeneous regression factor scores (HRFS) $\hat{\tilde{\rho}}_{rk}$ for the Target-rotated Big-Five factors E = Extraversion, ES = Emotional Stability, A = Agreeableness, C = Conscientiousness, and I = Intellect/Imagination based on 100 subsamples of $n$ = 150, $n$ = 600, and $n$ = 1,000 drawn randomly from the total sample of 696,854 cases; the error bars mark the 95% confidence interval. Non-overlapping confidence intervals show significant differences in determinacy coefficients between HRFS and RFS.

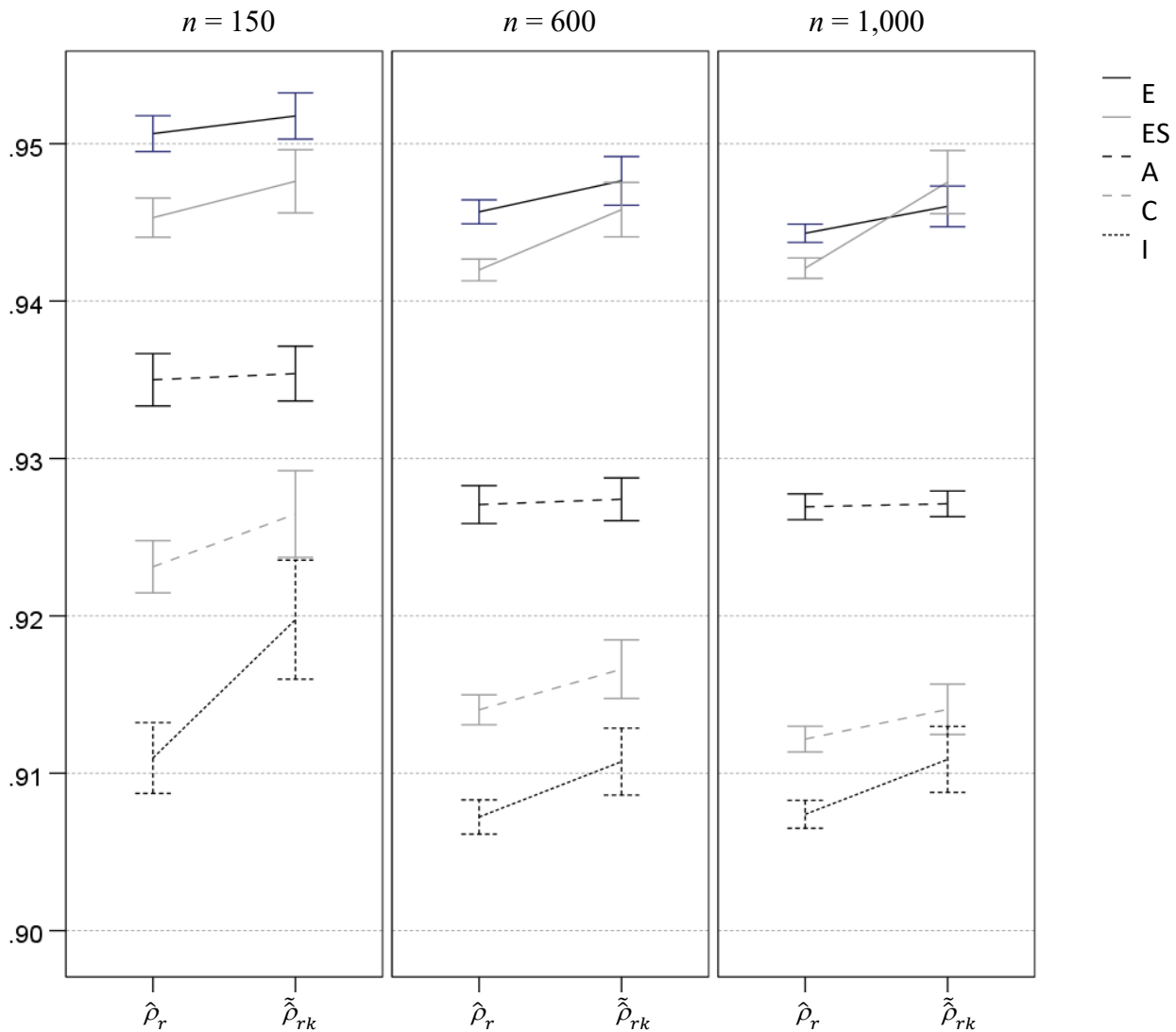

Figure 6. Mean parameter-based determinacy of regression factor scores (RFS) $\hat{\rho}_r$ and heterogeneous regression factor scores (HRFS) $\tilde{\hat{\rho}}_{rk}$ for the Varimax-rotated Big-Five factors E = Extraversion, ES = Emotional Stability, A = Agreeableness, C = Conscientiousness, and I = Intellect/Imagination based on 100 subsamples of $n$ = 150, $n$ = 600, and $n$ = 1,000 drawn randomly from the total sample of 696,854 cases; the error bars mark the 95% confidence interval. Non-overlapping confidence intervals show significant differences in determinacy coefficients between HRFS and RFS.

**Discussion**

Factor score predictors are supposed to estimate the latent value of individuals on the factor of interest. In both research and practical applications, it is important to use scores with maximum validity. The validity of factor score predictors is represented by their determinacy coefficients, that is, their correlation with the underlying factor. Molenaar et al. (2003) found that the determinacy coefficient of factor score predictors was reduced when heterogenous loadings which occur in in the population model were not specified in the factor model. Therefore, the present study investigated whether it is possible to improve the determinacy, and thus, the validity of the RFS by specifying heterogeneous loadings between individuals. We propose a method for the estimation of individual factor loadings and for the computation of HRFS. Moreover, we propose a binomial test to ascertain whether heterogeneous loadings are present. We suggest a two-step procedure: First, the binomial test should be conducted. If the binomial test for loading heterogeneity is significant, HRFS should be computed in the second step. Otherwise, RFS should be preferred.

The conditional computation of HRFS/RFS was compared to RFS by means of a simulation study based on population models with one and three factors and different degrees of loading heterogeneity. Population loading heterogeneity was quantified by the inter-individual standard deviation of loadings ($\sigma(\Lambda_{pop}) \in \{.00, .25, .50, .75\}$). Moreover, loading size, number of factors, number of variables per factor, and sample size were independent variables. Dependent variables of the simulation were factor score-based determinacies and parameter-based determinacies. Score-based determinacies were computed by means of the correlation of RFS and HRFS/RFS with the true factor scores used for the generation of the measured variables. Parameter-based determinacies were computed from the loading estimates and the inter-correlations of the measured variables. To eliminate the effect of rotation methods on results, loadings were based on principal axis factoring with subsequent orthogonal target-rotation. To investigate whether the results can be generalized, the simulation was repeated with Varimax-rotated factors for a subset of conditions.

Score-based determinacies and parameter-based determinacies were similar when there was no population loading heterogeneity ($\sigma(\Lambda_{pop}) = .00$). When population loading heterogeneity increased, parameter-based determinacy decreased more rapidly than score-based determinacy. As score-based determinacies represent a more direct measure of validity because they are based on the true factors, this indicates that the parameter-based determinacy underestimates the true determinacy under conditions of population loading heterogeneity. Moreover, with increasing population loading heterogeneity, parameter-based determinacy and score-based determinacy decreased more for RFS than for HRFS/RFS. This indicates that the latter can be recommended when the test of loading heterogeneity is significant, as we suggest by our two-step procedure. Larger loading heterogeneity resulted in larger differences of score-based determinacy between HRFS/RFS and RFS as well as larger differences of parameter-based determinacy between HRFS/RFS and RFS. Therefore, the difference of the parameter-based determinacies of HRFS/RFS and RFS can be used as an estimate for the difference of the score-based determinacies of HRFS/RFS and RFS. This is relevant because in empirical settings, the score-based determinacy cannot be computed and researchers need to rely on parameter-based determinacies. Moreover, the results of the simulation study were similar when Varimax-rotated loadings were computed instead of target-rotated loadings in the total sample, indicating that the results can be generalized to different rotation methods.

The empirical example based on random subsamples drawn from a large online sample of Big Five Factor Markers (Goldberg, 1992) revealed that for moderate sample sizes ($n = 600$) and for large sample sizes ($n = 1,000$), the parameter-based determinacy of the factor ES was larger for HRFS/RFS than for RFS. This result was found for target-rotated factors and for Varimax-rotated factors. As the result was only found for ES and for moderate to large samples, it is unlikely that it is due to sampling error. The results of the empirical example indicate that the computation of HRFS/RFS is also feasible for five factor models.

As a main limitation it should be noted that only orthogonal factors were investigated. The reason for this restriction is that – for correlated factors- individual differences of loadings may imply individual differences of factor inter-correlations. Although this concept could also be of interest, this would increase the number of estimated parameters, so that a related method would require very large sample sizes. Another limitation is that only the heterogeneity-based form of the RFS is considered. It might be of interest to use the individual loading estimates and the binomial test for loading heterogeneity for Bartlett factor scores and for correlation-preserving factor scores as well. Further limitations are that only Varimax-rotation was investigated as an orthogonal rotation method beyond target-rotation and that only population models with a simple structure of loadings were investigated. Future investigations could thus extend the approach of computing HRFS for different rotation methods, including oblique factor rotation and different estimation methods for factor scores.

**Conclusion**

The present research revealed that the negative effect of loading heterogeneity on the determinacy (validity) of RFS can be reduced when individual loadings are accounted for by computing HRFS. We propose a two-step procedure: First, a binomial test for loading heterogeneity is proposed. If the binomial test for loading heterogeneity is significant, we recommend the computation of HRFS. In our simulation study, he resulting conditional HRFS/RFS computation yielded an improvement of the factor score determinacy over the determinacy of the RFS in population models based on loading heterogeneity. An empirical example based on subsamples drawn randomly from a large Big Five Marker data set revealed that the ES factor may have substantial loading heterogeneity and that the determinacy may be improved by the conditional computation of HRFS/RFS.

**Acknowledgement**

This study was funded by the German Research Foundation (DFG), BE 2443/18-1.